# Automated Music Therapy for Anxiety and Depression Management in Older People (AMITY)

Malik Faizan[1], Dr. P.J. White[2], Dr. Indrakshi Dey[3].

(1) Walton Institute, Southeast Technological University (SETU)-Waterford, Ireland.
(2) DesignCORE-Southeast Technological University (SETU)-Carlow, Ireland.
(3) Walton Institute, Southeast Technological University (SETU)-Waterford, Ireland.

**ABSTRACT**: The onset of old age brings physiological and mental changes, with anxiety and depression being common mental disorders that can trigger other health issues and reduce lifespan. However, due to a global shortage of mental health professionals, combined with a growing population and limited awareness, these disorders often go undiagnosed. Music therapy offers a reliable method to address psychological, emotional, and cognitive needs. This paper presents an approach that monitors anxiety and depression symptoms in real time using low-complexity body sensors, followed by automated personalised music therapy, reducing the dependence on therapists and improving mental health care accessibility.

**Keywords:** Old age, Anxiety and Depression, Music therapy, Body Area Sensors, Mental Health Care

## 1. INTRODUCTION

The ageing process is inevitable, marking a significant phase of life where the body and mind undergo significant changes. The World Health Organization (WHO) defines the onset of old age as 60 years when an individual begins to experience natural physiological and mental changes. As age progresses, these changes become more evident, with older adults increasingly becoming more susceptible to physical ailments, cognitive decline, and mental health challenges. The recent COVID-19 pandemic has increased these risks, with older populations being particularly vulnerable to both physical and mental health issues. In recent years, the ageing population has grown rapidly, driven by two key factors: declining fertility rates and increasing life expectancy. According to the United Nations World Population Prospectus 2024, approximately 1 billion individuals were aged 60 or older in 2020, a figure that is projected to rise to 1.4 billion by 2030. By 2050, the number of people aged 60 and above will double, reaching an estimated 2.1 billion globally. In these statistics, individuals aged 80 or older are expected to triple from 2020 to 2050, soaring to around 426 million [1]. This unprecedented statistical shift presents opportunities and significant challenges for societies worldwide, particularly regarding healthcare, economic stability, and social support systems.

Despite the increase in older adults, a significant number continue to make important contributions to society. Advanced in age, many elders sustain their involvement in community activities, extending support to family members, dedicating their time to volunteer work, and sometimes remaining active in the workforce. These individuals frequently provide invaluable assistance with household management and offer younger generations a wealth of wisdom and experience. However, ageing also brings with it a host of challenges. Physical health naturally declines as the body ages and older adults are more prone to experiencing chronic illnesses, reduced mobility, and cognitive impairments. The collective impact of these physical challenges can lead to increased vulnerability to mental health conditions, such as depression and anxiety, even among those who remain physically healthy.

As people age, they often face multiple health issues simultaneously, which can significantly affect their mental well-being. Mental health disorders are particularly common in this age group, with approximately 14% of individuals aged 60 and above living with conditions such as depression and anxiety [2]. According to the Global Health Estimates (GHE) 2021, mental health conditions contribute to 10.6% of the total disability among older individuals [2]. These conditions diminish the quality of life and have broader societal impacts, burdening families, caregivers, and healthcare systems.

Depression and anxiety are among the most prevalent mental health conditions in older adults, and the statistics are alarming. A significant proportion of global suicides occur among individuals aged 60 or older, around 27.2%, underscoring the importance of the issue [3]. Physical health, life experiences, and the social environment influence mental health in old age. As individuals age, they often face hardships, such as chronic illnesses, loss of loved ones, retirement, and reduced income. These changes can lead to a decline in self-esteem and a sense of purpose, increasing the likelihood of psychological distress. Additionally, older adults are more likely to experience social isolation and loneliness, which can increase mental health conditions. Many older individuals may lose connections with friends and family due to death or relocation, and their social networks may shrink as they age. Ageism, or discrimination based on age, further adds to these challenges, leading to feelings of worthlessness and exclusion from society. The combined effects of these factors increase the risk of developing mental health disorders in older adults[3].

Despite the high prevalence of mental health disorders among the elderly, many of these conditions go unnoticed and untreated. Several barriers prevent older individuals from seeking help, and mental health issues often remain underreported. One of the primary reasons is the stigma associated with mental health. Older adults may hesitate to discuss their emotional and psychological struggles due to societal norms that discourage openness about mental health issues. There is a common belief in many cultures that mental illness is a sign of weakness or moral failure, which prevents older individuals from seeking help. This unwillingness is often accompanied by the fear of being judged by others, including family members and healthcare professionals. Another critical barrier is the lack of access to mental health services, particularly in low and middle-income countries [4]. Even developed countries are facing a shortage of mental health professionals capable of addressing the needs of the elderly. The rapid growth in the number of older adults has surpassed the availability of trained professionals. Furthermore, there is a lack of public awareness about mental health in older age, which contributes to delayed diagnosis and treatment [5]. Without timely intervention, mental health conditions can worsen, leading to more severe outcomes such as cognitive decline, physical health deterioration, and even suicide.

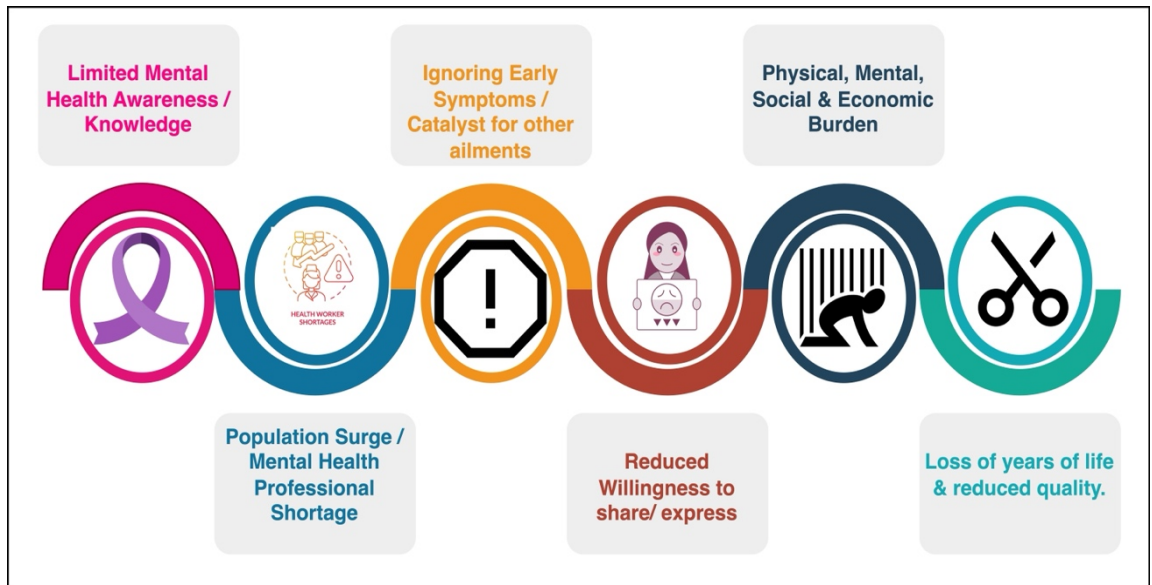

Figure 1: Issue related to mental disorders.

To overcome the growing mental health crisis among older adults, it is essential to propose a solution that addresses the challenges. Traditional methods of mental health treatment, such as psychotherapy and medication, may not always be accessible for older individuals, particularly those with limited mobility or those living in remote areas. Therefore, alternative approaches that prioritise accessibility, privacy, and ease of use are crucial for addressing mental health issues. One such solution is integrating technology into mental health care for older adults. The paper proposes an approach that utilises low-complexity, non-invasive body area photoplethysmography (PPG) sensors to monitor the symptoms of anxiety and depression in real time. The data collected by the sensors would be transmitted to a cloud-based computing module, where it would be analysed and classified without causing discomfort. This approach would not only provide real-time monitoring of mental health symptoms but also ensure that older adults receive timely interventions. The cloud-based module would offer easy accessibility, enabling family members and healthcare providers to monitor the mental health of elderly individuals remotely. Privacy would be a key priority, ensuring that personal data is protected while providing valuable information regarding an individual's mental health status. In addition to real-time monitoring, the proposed system would incorporate personalised music therapy as a form of treatment for mental health disorders. By offering a user-centric delivery system for streaming personalised music, the system would provide therapeutic interventions that do not rely on the constant presence of mental health professionals. This method would reduce the need for traditional approaches, such as sedatives or therapy sessions, which may not always be practical for older adults.

## 2. BACKGROUND

The common mental disorders are either depression disorders or anxiety disorders. Depressive disorders are characterised by sudden mood changes and are often categorised as dysregulation disorders. Common symptoms include sadness, emptiness, hopelessness, worthlessness, despondency, irritability, and changes in somatic and cognitive functions. In older adults, depression often presents with distinct clinical characteristics compared to younger individuals [6]. This is mainly due to the increased likelihood of comorbidities, differences in the underlying causes, and variations in symptom expression. The cause of depression in the elderly is more complex, often influenced by age-related changes in the brain, as well as the presence of neurodegenerative and cardiovascular diseases, which can contribute to the development of depression in later life [6,7]. On the other hand, Anxiety disorder is characterised by excessive fear, anxiety, behavioural disturbances and anticipation of future threats. These disorders often persist over time and are directly related to the age or timing of the triggering incident. Anxiety and fear share common features, both closely associated with the sympathetic nervous system's fight-or-flight response. The lifetime prevalence of anxiety disorders in older adults is significantly higher. Studies by Lenze et al. show that approximately 35% of elderly individuals have been diagnosed with an anxiety disorder at some point in their lives, with 24% having received a recent diagnosis [8]. Furthermore, comorbidity between depression and anxiety can be as high as 48.3% [9]. The risk of mortality due to these disorders has increased in older adults [10], with suicide rates being particularly high among older men [11].

Research shows that the risk of suicide increases threefold in individuals aged 75 or older compared to younger age groups [12]. Anxiety disorders are also linked to increased suicides [8]. The elderly population, particularly during the COVID-19 pandemic, experienced heightened levels of loneliness, distress, and exhaustion, with anxiety contributing to the development of depression, as noted by Bergdahl and Bergdahl [13]. Additionally, factors such as the loss of a partner and declining health increase the risk of depression among the elderly. On the other hand, social capital resources from social networks can be a protective factor, contributing to mental well-being in this age group [14]. The 2022 *Mental Health* European House headway report highlighted the growing prevalence of mental health disorders across Europe. Approximately one in six individuals in Europe is affected by mental health conditions, including anxiety, depression, and bipolar disorder [15]. According to the World Health Organization (WHO), the COVID-19 pandemic led to a 25% rise in mental health disorders, particularly anxiety and depression [1]. The impact of these disorders varies significantly across Europe, with 16.9 million living lives with disability due to mental health disorders even before the pandemic. Depression is the fourth leading cause of disability in the region, followed by anxiety and schizophrenia [2]. Mental and behavioural disorders account for approximately 4% of all deaths in Europe, with particularly high mortality rates among those with severe mental illness. Ireland reports around 388 annual deaths, with the cause being mental health issues[15]. Women and individuals over 65 are especially at higher risk, with these conditions having a significant impact on their quality of life and mortality.

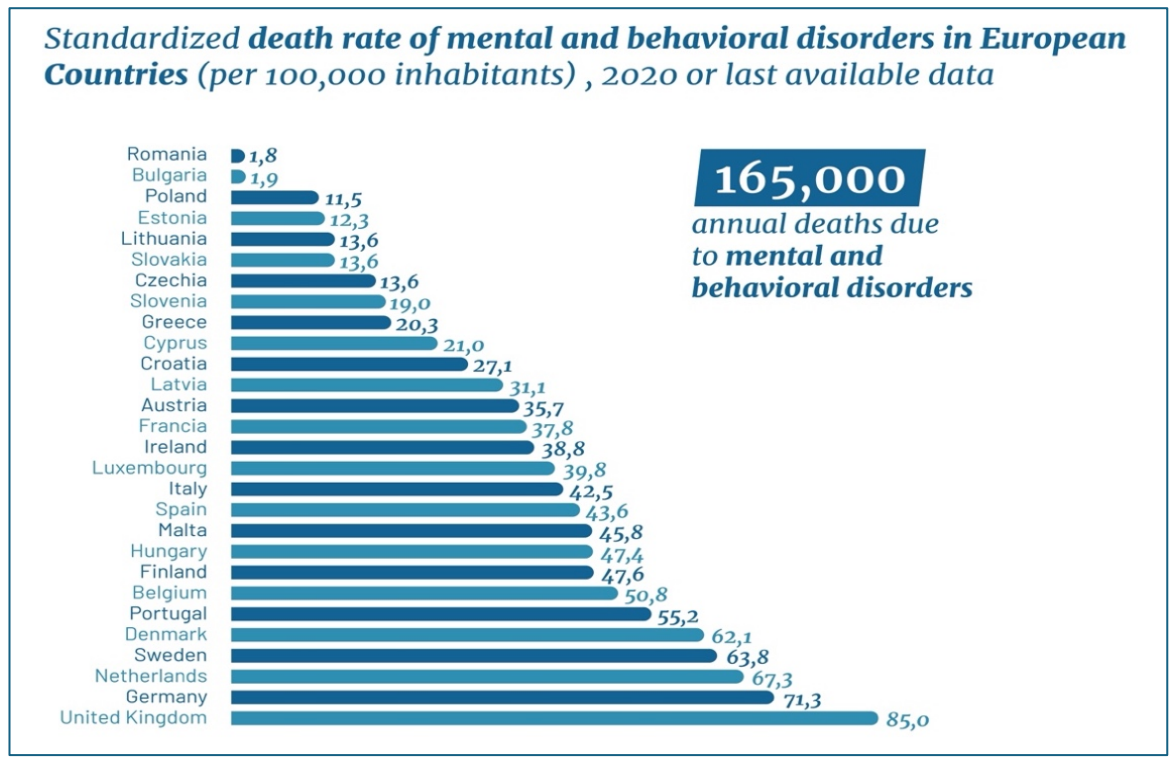

Figure 2: Prevalence of Mental and Behavioural disorders in Europe [15]

Although numerous studies have been conducted on mental disorders and their prevalence in old age, individuals still heavily rely on traditional methods that have limitations in terms of accessibility and capacity and require the constant presence of therapists or mental health professionals. The use of medication such as sedatives or substances containing serotonin reuptake inhibitors (SRI) has been the primary solution for addressing mental disorders. While these treatments have been successful over the years, they have also led to increased dependence on drugs among individuals of numerous age groups.

3. **METHODOLOGY**

Anxiety and depression represent the most prevalent mental disorders worldwide and often co-occur, causing a sudden shift from a calm to an anxious state in affected individuals. Long-term frequency anxiety disorders lead to the development of depressive disorders. Anxiety and depression disorders share common somatic symptoms across all age groups, such as

- Palpitations/ pounding of heart/ increased heart rate
- Breathlessness/ elevated respiratory rate
- Increased blood pressure.
- Irregular heart rate variability.
- Disrupted sleep cycles.
- Vertigo/loss of consciousness, sweating, trembling, shaking.
- Dizziness, lightheadedness or faintness, fatigue.
- Paranesthesia (numbness or tingling sensation).
- Agitated psychomotor/ restlessness.

Most of these somatic symptoms can be directly monitored using a body area photoplethysmography (PPG) sensor in the form of a wristwatch. The PPG sensor can continuously monitor physiological parameters, which include heart rate, blood pressure, respiratory rate, sleep patterns, and activity levels, in real time without causing discomfort. The data recorded by the sensor is transmitted directly to a cloud or personalised device for collection and processing. Certain data analysis techniques are applied to classify and correlate the data, identifying parameters associated with symptoms of mental disorders, particularly anxiety and depression. When the parameters surpass a predefined threshold corresponding to the symptoms of anxiety and depression, an activation function is triggered. This activation function is transmitted to a user-centric delivery system such as a headset or earpiece to stream personalised music as therapy onto an individual's ears. Personalised/ tailored music has a significant effect on modulating brain waves. It activates the parasympathetic response, causing the normalising of physiological parameters and gradually transitioning an individual from an anxious state to a state of calm. Using music to stimulate parasympathetic responses also promotes positive cognitive ideation.

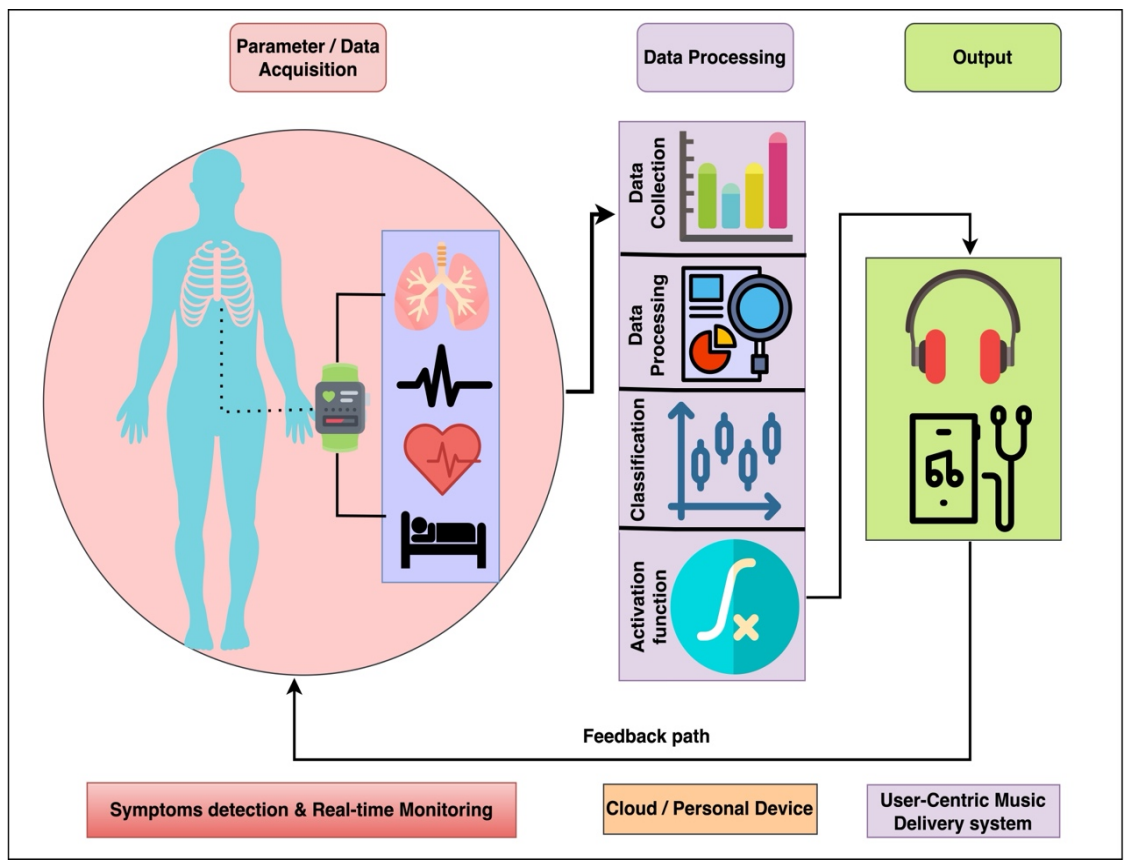

Figure 3: Block Diagram Representation.

The entire process functions within a feedback loop, continuously recording physiological parameters. If these physiological parameters drop to levels below the threshold, the music therapy automatically stops. This innovative approach offers accessible, swift, and user-driven support for mental health, empowering individuals, particularly older adults, to take proactive and autonomous measures in managing their mental well-being.

## 4. POSSIBLE OUTCOME

The symptoms of anxiety and depression typically present at the onset of the disorder without any identifiable initial stressors. The disorder induces a transition in an individual's state from calm to a state of anxiety. The autonomic nervous system is responsible for governing the body's unconscious activities. Personalised or tailored music therapy significantly improves physical and mental health by stimulating the receptors of the parasympathetic nervous system. The parasympathetic nervous system facilitates the body's recovery in a stressful situation by reducing heart and respiratory rates. Parasympathetic receptors are located in the submandibular region of the brain and as muscarinic receptors in the heart. These receptors regulate brain waves to improve an individual's emotional well-being [16]. Music therapy significantly influences an individual's physical and psychological recovery by triggering the parasympathetic response caused by the modulation of brain waves. It enhances cognitive recovery, mitigates negative emotions, and aids in restoring individuals to a state of calm [17].

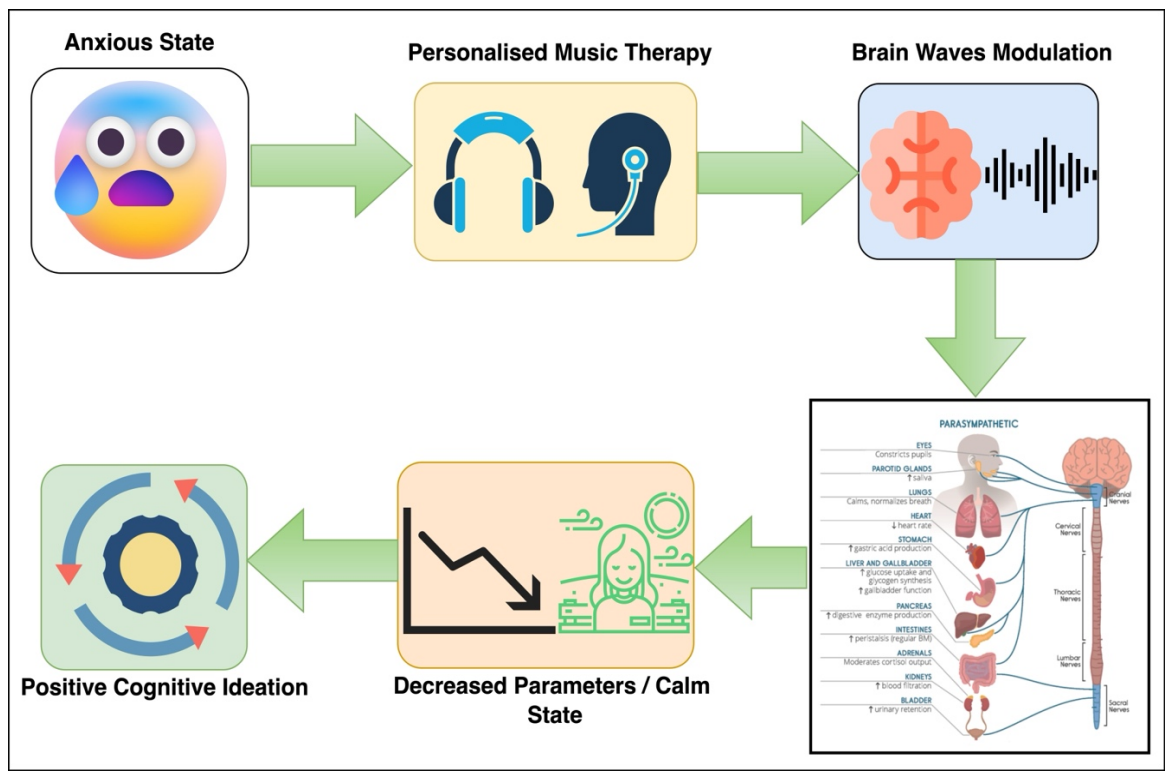

Figure 4: Schematic effect of music therapy in regulating mental well-being.

Music directly affects the limbic system and brainstem, influencing physical and physiological functions. Its melody has the capacity to uplift the spirit, stimulate cognitive processes, promote typical behavioural responses, enhance emergency response capabilities, particularly during panic attacks and facilitate the utilisation of the patient's inherent well-being abilities [18].

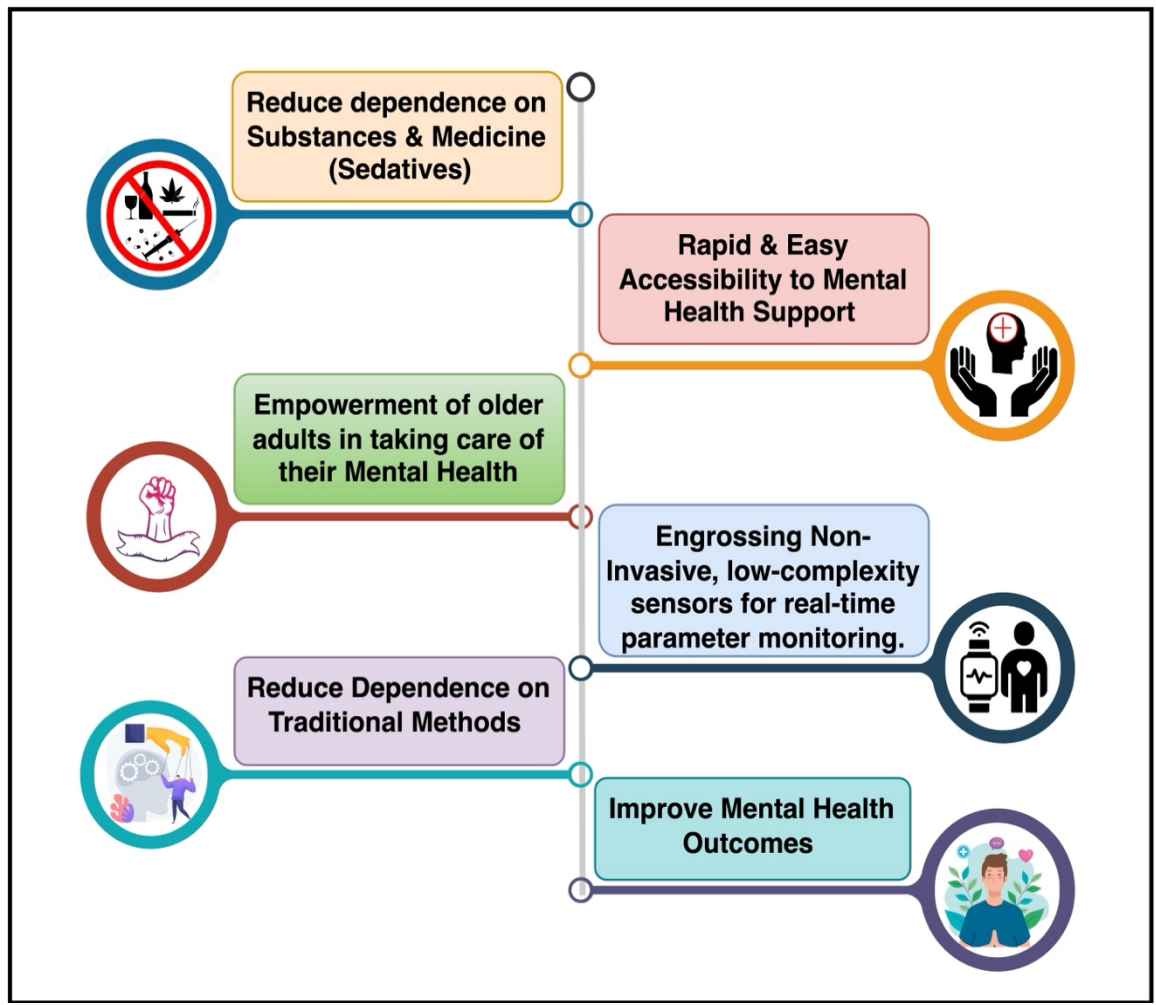

Figure 5: Impact of automated music therapy.

5. **CONCLUSION**

The integration of sensor technology with cloud computing and adaptive music therapy is designed to empower elderly individuals to actively and independently manage their mental well-being. The ability of personalised or tailored music to influence brain waves and initiate the parasympathetic response plays a crucial role in regulating patients' physiological parameters towards normalisation. Thus facilitating a transition from an anxious state to a calm state and fostering positive cognitive ideation. Consequently, this creates a potential for reducing the dependence on medication and enhancing overall physical and mental well-being.

**Acknowledgement**

I gratefully acknowledge the support of South East Technological University-Waterford, Ireland, in funding this project. The research is progressing in its early stages at Walton Institute. The project reference number is WD_2023_46_WSCH.

# References

1. "World Population Prospects 2024 Summary of Results Ten key messages," 2024. https://population.un.org/wpp/Publications/Files/WPP2024_Key-Messages.pdf
2. Institute for Health Metrics and Evaluation, "Global Health Data Exchange (GHDx)," *Institute for Health Metrics and Evaluation*, 2021. https://vizhub.healthdata.org/gbd-results.
3. J. Proudfoot, G. Parker, D. Hadzi Pavlovic, V. Manicavasagar, E. Adler, and A. Whitton, "Community Attitudes to the Appropriation of Mobile Phones for Monitoring and Managing Depression, Anxiety, and Stress," *Journal of Medical Internet Research*, vol. 12, no. 5, p. e64, Dec. 2010, doi: https://doi.org/10.2196/jmir.1475.
4. A. S. Young, R. Klap, C. D. Sherbourne, and K. B. Wells, "The Quality of Care for Depressive and Anxiety Disorders in the United States," *Archives of General Psychiatry*, vol. 58, no. 1, p. 55, Jan. 2001, doi: https://doi.org/10.1001/archpsyc.58.1.55.
5. E. Brohan and G. Thornicroft, "Stigma and discrimination of mental health problems: workplace implications," *Occupational Medicine*, vol. 60, no. 6, pp. 414–415, Aug. 2010, doi: https://doi.org/10.1093/occmed/kqq048.
6. C. G. Gottfries, "Is there a difference between elderly and younger patients with regard to the symptomatology and aetiology of depression," *International Clinical Psychopharmacology*, vol. 13, no. 5, p. S13, Sep. 1998, Available: https://journals.lww.com/intclinpsychopharm/Abstract/1998/09005/Is_there_a_difference_between_elderly_and_younger.4.aspx
7. C.-G. Gottfries, "Late life depression," *European Archives of Psychiatry and Clinical Neuroscience*, vol. 251, no. S2, pp. 57–61, Jun. 2001, doi: https://doi.org/10.1007/bf03035129.
8. E. J. Lenze, "Comorbid Anxiety Disorders in Depressed Elderly Patients," *American Journal of Psychiatry*, vol. 157, no. 5, pp. 722–728, May 2000, doi: https://doi.org/10.1176/appi.ajp.157.5.722.
9. R. Johansson, P. Carlbring, Å. Heedman, B. Paxling, and G. Andersson, "Depression, anxiety and their comorbidity in the Swedish general population: point prevalence and the effect on health-related quality of life," *PeerJ*, vol. 1, p. e98, Jul. 2013, doi: https://doi.org/10.7717/peerj.98
10. M. A. Rapp, D. Gerstorf, H. Helmchen, and J. Smith, "Depression Predicts Mortality in the Young Old, but Not in the Oldest Old: Results From the Berlin Aging Study," *The American Journal of Geriatric Psychiatry*, vol. 16, no. 10, pp. 844–852, Oct. 2008, doi: https://doi.org/10.1097/jgp.0b013e31818254eb.
11. D. De Leo, "Late-life suicide in an aging world," *Nature Aging*, vol. 2, no. 1, pp. 7–12, Jan. 2022, doi: https://doi.org/10.1038/s43587-021-00160-1.
12. A. Ojagbemi, B. Oladeji, T. Abiona, and O. Gureje, "Suicidal behaviour in old age - results from the Ibadan study of ageing," *BMC Psychiatry*, vol. 13, no. 1, Mar. 2013, doi: https://doi.org/10.1186/1471-244x-13-80.
13. J. Bergdahl and M. Bergdahl, "Perceived stress in adults: prevalence and association of depression, anxiety and medication in a Swedish population," *Stress and Health*, vol. 18, no. 5, pp. 235–241, 2002, doi: https://doi.org/10.1002/smi.946.
14. F. Nyqvist, A. K. Forsman, G. Giuntoli, and M. Cattan, "Social capital as a resource for mental well-being in older people: A systematic review," *Aging & Mental Health*, vol. 17, no. 4, pp. 394–410, May 2013, doi: https://doi.org/10.1080/13607863.2012.742490.
15. "The incidence of mental disorders in European countries," *www.angelinipharma.com*. https://www.angelinipharma.com/our-responsibility/headway-a-new-roadmap-in-mental-health/the-incidence-of-mental-disorders-in-european-countries.
16. S. Jentschke and S. Koelsch, "Gehirn, Musik, Plastizität und Entwicklung," *VS Verlag für Sozialwissenschaften eBooks*, pp. 51–70, Jan. 2006, doi: https://doi.org/10.1007/978-3-531-90607-2_5.
17. Y. Cha, Y. Kim, S. Hwang, and Y. Chung, "Intensive gait training with rhythmic auditory stimulation in individuals with chronic hemiparetic stroke: A pilot randomised controlled study," *NeuroRehabilitation*, vol. 35, no. 4, pp. 681–688, 2014, doi: https://doi.org/10.3233/nre-141182.

18. T. Sarkamo *et al.*, "Music listening enhances cognitive recovery and mood after middle cerebral artery stroke," *Brain*, vol. 131, no. 3, pp. 866–876, Feb. 2008, doi: https://doi.org/10.1093/brain/awn013.